\documentclass[a4paper,11pt]{article}
\pdfoutput=1
\usepackage{jheppub}

\usepackage{bm,wasysym}

\def\slash#1{\setbox0=\hbox{$#1$}\dimen0=\wd0
      \setbox1=\hbox{/} \dimen1=\wd1 \ifdim\dimen0>\dimen1
      \rlap{\hbox to \dimen0{\hfil/\hfil}} #1                        \else
      \rlap{\hbox to \dimen1{\hfil$#1$\hfil}}
      /   \fi}

\newcommand{\lsim}{
\mathrel{\hbox{\rlap{\hbox{\lower4pt\hbox{$\sim$}}}\hbox{$<$}}}}

\newcommand{\gsim}{
\mathrel{\hbox{\rlap{\hbox{\lower4pt\hbox{$\sim$}}}\hbox{$>$}}}}

\allowdisplaybreaks[2]

\newcommand{\ord}{{\cal O}}

\title{Repercussions of Flavour Symmetry Breaking  \\[0.3em] on  CP Violation in $D$-Meson Decays}

\author[a]{Thorsten Feldmann,}
\author[a]{Soumitra Nandi,}
\author[b]{and Amarjit Soni}

\affiliation[a]{Theoretische Elementarteilchenphysik, 
Naturwissenschaftlich Technische Fakult\"at, \\
Universit\"at Siegen, 57068 Siegen, Germany}

\affiliation[b]{Physics Department, Theory Group, Brookhaven National Laboratory, \\  Upton, NY 11973, USA}

\emailAdd{thorsten.feldmann@uni-siegen.de}

\emailAdd{nandi@physik.uni-siegen.de}

\emailAdd{soni@bnl.gov}

\abstract{We investigate to what extent the recently measured value for
a non-vanishing direct CP asymmetry in $D^0 \to K^+K^-$ and $D^0 \to \pi^+\pi^-$
decays can be accommodated in the Standard Model (SM) or extensions with a
constrained flavour sector, 
for instance from a sequential $4^{\rm th}$ generation of quarks (4G).
From the comparison with $D^0 \to K^- \pi^+$ branching ratios, we establish
large U-spin symmetry ($d \leftrightarrow s$) breaking effects 
with large strong phases between different interfering amplitudes.
On the basis of conservative estimates on amplitude ratios ---
which are supported by an analysis of the breaking 
of a $c\leftrightarrow u$ symmetry in non-leptonic $B^0$ decays ---
we find that, in the SM, direct CP asymmetries in the $\pi^+ \pi^-$ or $K^+ K^-$ modes
(or in their difference) of the order of several per mille are still plausible.
Due to the constraints on the new CP phases in the 4G model, only moderate effects 
compared to the SM estimates are possible.
We suggest CP studies at LHCb as well as at (Super)B-factories
of several distinctive modes, such as $D^+ \to \bar K^{(*)0}\pi^+, \phi \pi^+$ 
and $D_s \to K^{(*)0} \pi^+,\phi \pi^+(K^+)$ etc., which should shed more light on the short- 
and long-distance issues underlying CP violation in non-leptonic D-meson decays.
}


\note{Preprint SI-HEP-2012-03,  \today}

\begin{document}

\maketitle


\section{Motivation}

CP-violation studies in the charm system are important for a variety of reasons.  First, 
in the Standard Model (SM), only small asymmetries are expected, see e.g.~\cite{Browder_RMP,Artuso:2008vf,Antonelli:2009ws,Buchalla:2008jp}
and references therein.  On the 
other hand, in many models of new physics (NP), the top quark is sensitive to non-standard effects, 
and there are flavour-changing interactions which couple the charm and the top quark.  
Flavour models based on generic ideas of warped extra dimensions represent a very interesting  
example of this type; especially so as large $\ord(1)$ CP-odd phases may accompany the new 
interactions~\cite{warp}.  This is also the case in many models with an extended Higgs 
sector~\cite{T2HDM}.  Even in a simple extension of the SM with a fourth generation of fermions,
significant differences in CP asymmetries in the charm sector may arise as compared to the SM~\cite{AJB_etal,SNAS,Bobrowski:2010xg}.
Furthermore, the  charm-quark mass ($\approx$1.2~GeV) is not that heavy;  consequently large (CP-even) 
rescattering phases can be present in $D$-meson decays which, together with CP-odd phases from the SM and/or NP sector,
can lead to sizeable direct CP asymmetries.  CP studies via $D^0$-$\bar D^0$ mixing are also highly motivated as
the $D^0$ is a unique bound state of charge~$2/3$ quarks and is thus sensitive to NP affecting the top quark.


In this article, we will be mainly concerned with CP violation in non-leptonic $D^0$ decays.
The time-integrated CP asymmetry $A_{\rm CP}(f)$ for a given final state $f$ is defined as 
\begin{equation}
A_{\rm CP}(f) = \frac{\Gamma(D^0\to f) - \Gamma({\bar D}^0\to f)}{\Gamma(D^0\to f) + \Gamma({\bar D}^0\to f)}, 
\end{equation}
which (to first approximation) may be decomposed as (see e.g.\ \cite{Grossman:2006jg})
\begin{equation}
A_{\rm CP}(f) = A_{\rm CP}^{\rm dir} + A_{\rm CP}^{\rm ind} + A_{\rm CP}^{\rm mix}.
\label{acpf}
\end{equation} 
Here, $A_{\rm CP}^{\rm dir}(f)$ is the direct CP asymmetry in the decay $D^0\to f$, 
$A_{\rm CP}^{\rm mix}$ is the CP asymmetry from $D^0$-$\bar D^0$ mixing, and  $A_{\rm CP}^{\rm ind}$ stems
from the interference of mixing and decay.
%
%
Recent results from the LHCb experiment~\cite{LHCb} on CP asymmetries in $D^0$ decays,
\begin{equation}
\Delta A_{\rm CP}^{\rm dir} \equiv A_{\rm CP}^{\rm dir}(K^+K^-) - A_{\rm CP}^{\rm dir}(\pi^+\pi^-) = - \left( 0.82 \pm 0.21 \pm 0.11 \right)\% \,,
\label{lhcb}
\end{equation}
indicate a 3.5$\sigma$ deviation from 0, with a large amount of experimental systematics cancelling in
the considered difference of decay modes. We remind the reader that, in the $SU(3)_F$ limit for light quarks
\cite{Grossman:2006jg}, 
the direct CP asymmetries in the individual channels are equal in magnitude but opposite in sign,
$A_{\rm CP}^{\rm dir}(K^+K^-) \simeq - A_{\rm CP}^{\rm dir}(\pi^+\pi^-)$.
Assuming that the effects of $A_{\rm CP}^{\rm mix}$ drop out in the difference, 
and that the effect of $A_{\rm CP}^{\rm ind}$ is small, 
this points to a relatively large value for $|A_{\rm CP}^{\rm dir}|$
for both decay modes. Not surprisingly, the LHCb result has renewed the interest in CP violation in charm physics,
and a number of articles, addressing the interpretation of the data within the SM or NP, have
appeared since (see e.g.\ \cite{arXiv:1103.5785,arXiv:1110.2862,arXiv:1111.4987,arXiv:1111.5000,Hochberg:2011ru,Rozanov:2011gj,Pirtskhalava:2011va,Cheng:2012wr,Bhattacharya:2012ah,Giudice:2012qq,Altmannshofer:2012ur,Jung:2012ip}). 
At about the same time,  the CDF collaboration has reported their result from an independent measurement
\cite{arXiv:1111.5023}, which also provides numbers for the individual CP asymmetries in
the $(K^+K^-)$ and $(\pi^+\pi^-)$ channel, 
\begin{align}
A_{\rm CP}(K^+K^-) &= (-0.24 \pm 0.22 \pm 0.09)\% \,, \cr
A_{\rm CP}(\pi^+\pi^-) &= (+0.22 \pm 0.24 \pm 0.11) \% \,,
\label{cdfd0}
\end{align}
which is consistent with the above LHCb result. 
Both measurements dominate the world average provided by the HFAG collaboration \cite{hfag},
\begin{align}
\Delta A_{\rm CP}^{\rm dir} = (-0.645 \pm 0.180 )\% \,. 
\label{hfag_avg}                                                    
\end{align}
%


Although a direct CP asymmetry of the order of a few per-mille seems to be a small number,
the SM expectations for $\Delta A_{\rm CP}^{\rm dir}$ in $D$ decays are usually even smaller,
because the only source of observable CP violation comes from the interference 
with a sub-leading amplitude which involves a relative suppression factor involving 4 powers of
the Cabibbo angle. A precise quantitative estimate, however, is difficult because
standard approximation methods known from $B$-meson or kaon physics do not seem to work 
sufficiently well in the charm sector, since the charm-quark mass is neither much smaller
nor much larger than the typical hadronic scales.
Without strong theoretical prejudices about non-perturbative hadronic dynamics, we are therefore constrained to 
semi-quantitative analyses. The related questions that we are going to explore in this paper 
are:
\begin{itemize}
 \item What can be learned from U-spin relations between down- and strange-quarks, and how large is the effect of U-spin breaking?
 \item Do we have phenomenological evidence for large strong phases in $D^0 \to P^+P^-$ decays, and how does this compare
       with the situation in non-leptonic $B^0$-meson decays?
 \item How large do sub-leading amplitudes in the SM have to be in order to explain the observed $\Delta A_{\rm CP}^{\rm dir}$?
 \item To what extent can NP models with constrained flavour sector, like for instance a model with 4 fermion
        generations, enhance the theoretical predictions for  $\Delta A_{\rm CP}^{\rm dir}$ as compared to the SM?
 \item Can we identify other $D$-decay modes which should exhibit similar patterns of CP violation?  
\end{itemize}

\section{$U$-Spin in Non-Leptonic $D^0 \to P^+P^-$ Decays}

Assuming $SU(3)$ flavour symmetry for the strong interactions of light quarks,
the contributions to the amplitudes for non-leptonic $D$-meson decays
can be related. 
For our purposes, it will be sufficient to focus on a sub-group, U-spin symmetry,
which acts on U-spin doublets $(d,s)$ and which have been frequently used 
to analyse weak non-leptonic meson decays, see for instance
 \cite{Fleischer:1999pa,Gronau:2000zy,Soni:2006vi,Jung:2009pb}.

Starting point is the weak effective Hamiltonian for $c \to u q \bar q'$ transitions ($q,q' = u,d,s$)
which can be decomposed into a $U=0$ and $U=1$ part, as follows
\begin{align}
 H_{\rm eff}(c \to u s\bar d) &=  -\left(V_{cs}^* V_{ud}\right)  H_{U=1}^{(U_3=-1)} \,, 
\nonumber \\[0.25em]
 H_{\rm eff}(c \to u q\bar q) &=  \left( \frac{V_{cd}^* V_{ud} - V_{cs}^* V_{us}}{\sqrt 2} \right)  H_{U=1}^{(U_3=0)} + \left( V_{cd}^* V_{ud} + V_{cs}^* V_{us} \right)  H_{U=0} \,,
\nonumber \\[0.25em]
 H_{\rm eff}(c \to u d\bar s) &=  \left( V_{cd}^* V_{us} \right) H_{U=1}^{(U_3=+1)} \,. 
\end{align}
Here $H_{U=1}$ only contains the current-current operators from tree-level $W$-boson exchange, while $H_{U=0}$ also receives contributions 
from strong and electro-weak penguin operators. The important property to notice is that the U-spin singlet or triplet terms are multiplied 
by a single combination of Cabibbo-Kobayashi-Maskawa (CKM) elements.
In the exact U-spin limit, the amplitudes for the related $D^0$-meson decays into two light charged pseudoscalar mesons\footnote{We are restricting
ourselves to charged mesons (with $U=1/2$) here, because in this way we only generate $U=0$ and $U=1$ final states which require a minimal set
of independent amplitudes, once we allow for arbitrary U-spin breaking, see below.}
are thus described by only two independent complex amplitudes, which we denote by $A_{U=0}$ and $B_{U=1}$, 
\begin{align}
 \mbox{U-spin limit:} &&{\cal A}[D^0 \to K^- \pi^+] &= 2 \,V_{cs}^* V_{ud} \, B_{U=1} \,,
\nonumber \\[0.25em]
&&
{\cal A}[D^0 \to \pi^+ \pi^-] &= (\lambda_d + \lambda_s) \, A_{U=0} + (\lambda_d -\lambda_s) \, B_{U=1} \,,
\nonumber \\[0.25em]
&& {\cal A}[D^0 \to K^+ K^-] &= (\lambda_d + \lambda_s) \, A_{U=0} - (\lambda_d -\lambda_s) \, B_{U=1} \,,
\nonumber \\[0.25em]
&& {\cal A}[D^0 \to K^+ \pi^-] &= 2 \,V_{cd}^* V_{us} \, B_{U=1} \,,
\end{align}
where we have defined
$ \lambda_d \equiv V_{cd}^* V_{ud}$ and $\lambda_s \equiv V_{cs}^* V_{us}$.
From the Wolfenstein expansion of the CKM elements in powers of $\lambda = \sin\theta_C \sim \ord(0.2)$, we infer that 
the decay $D^0 \to K^- \pi^+$ is Cabibbo allowed (CA), the decays $D^0 \to \pi^+\pi^-, K^+K^-$ are singly-Cabibbo-suppressed (SCS)
with $(\lambda_d-\lambda_s) \sim \ord(\lambda)$, while the decay $D^0 \to K^+\pi^-$ is double-Cabibbo-suppressed (DCS).
Furthermore, the contribution of the $U=0$ amplitude to the decays $D^0 \to \pi^+\pi^-$ and $D^0 \to K^+K^-$ is suppressed
by another 4 powers of $\lambda$, since $\lambda_d+\lambda_s \sim \ord(\lambda^5)$,
and therefore the decay rates should be equal, while the CP asymmetries from the interference
of $A_{U=0}$ and $B_{U=1}$ should be tiny, as already mentioned above. 
However, the measured branching ratios (BRs) for the CA, SCS, DCS modes (experimental numbers
are taken from \cite{hfag} or \cite{pdg}, see also references to the original experiments
therein)
\begin{align}
{\rm BR}[D^0 \to K^-\pi^+] &= (3.949 \pm 0.023 \pm 0.040 \pm 0.025)\% 
\,,
\nonumber\\[0.25em]
{\rm BR}[D^0 \to \pi^+\pi^-] &= (0.1425 \pm 0.0019 \pm 0.0018 \pm 0.0014)\% 
\,, \
\nonumber \\[0.25em]
{\rm BR}[D^0 \to K^+K^-] &= (0.3941 \pm 0.0038 \pm 0.0050 \pm 0.0024)\% 
\,, \
\nonumber \\[0.25em]
\frac{{\rm BR}[D^0 \to K^+\pi^-]}{
{\rm BR}[D^0 \to K^-\pi^+]} &= (0.331 \pm 0.008)\% \,,
\end{align}
do not follow the pattern expected from U-spin symmetry. 
Actually,  correcting for phase-space effects and CKM factors, the following observables
\begin{align}
{\rm obs}_1 & \equiv \frac{{\rm BR}[D^0 \to K^+K^-]/|\vec p_K|}{{\rm BR}[D^0 \to \pi^+\pi^-]/|\vec p_\pi|}
\simeq 3.22 \pm 0.09 \label{obs1}
\\[0.25em]
{\rm obs}_2 & \equiv 
\frac{{\rm Br}[D^0 \to K^- \pi^+ ]/|\vec p_{\pi K}|}{{\rm Br}[D^0 \to K^+ K^-]/|\vec p_K|} \, \lambda^2 \simeq 0.47 \pm 0.01
\,,
\label{obs2} 
\\[0.25em]
{\rm obs}_3 & 
\equiv 
\frac{{\rm Br}[D^0 \to K^+ \pi^- ]}{{\rm Br}[D^0 \to K^- \pi^+]}\, \lambda^{-4} \simeq  1.28 \pm 0.03
\,,
\label{obs3}
\end{align}
deviate significantly from unity. In principle, this can be due to:
U-spin violation originating from $m_s \neq m_d$ in the SM, and/or a drastic enhancement of the hadronic matrix elements 
of penguin operators in $A_{U=0}$ either from long-distance dynamics in the SM or short-distance effects 
from sizeable NP. In the latter case, however, it can easily be seen that the enhancement 
of order $\lambda^{-4} \sim \ord(600)$ necessary to reproduce the pattern of BRs, at the same time clashes
with the (still small) values for the direct CP asymmetries, which in such a case would be enhanced by
the same factor, unless the strong phases of $A_{U=0}$ and $B_{U=1}$ were fine-tuned.
Actually, we will find below that the simultaneous fit to ${\rm obs}_1$ and ${\rm obs}_2$ clearly fixes
the strong-phase differences of the relevant interfering amplitudes in a scenario with broken U-spin to be large.

In the following, we are thus allowing for generic U-spin violation in the above decay amplitudes.
In order to explain the measured BRs, we would have to accept U-spin violating effects as large as
50-60\%\ on the amplitude level. In the naive factorization approach, such factors can largely 
be explained by the difference in decay constants ($f_K/f_\pi$) and hadronic form factors
($F(D\to K)/F(D\to \pi)$) each of which amounts to $\ord(20\%)$ corrections to the U-spin limit
(recent lattice results\footnote{We thank Ruth Van de Water for discussions pertaining to these.}
on these quantities can, for instance, be found in 
\cite{lattice_averages,HPQCD_11}). As already mentioned, the observed BRs also require a 
significant strong-phase difference which, together with the remaining amount of U-spin
violation in the magnitude of the decay amplitudes, points towards essential non-factorizable
long-distance effects in the hadronic matrix elements. 
Again, this comes as no surprise, as similar conclusions have been drawn from the analysis
of non-leptonic $B$-meson decays 
(see e.g.\ \cite{ANS98,Fleischer:1998bb,Deshpande:1998xq,Rosner:1999zm,Buras:2000gc,Buras:2004ub,Feldmann:2004mg,Cheng:2004ru,JLR,Feldmann:2008fb,HYC}
for an incomplete list of references),
although the importance of non-factorizable contributions is appreciably less dramatic in $B$-meson decays due to
the fact that the $1/m_b$ expansion  there is more efficient than 
a $1/m_c$ expansion in $D$-meson decays.

\subsection{Including U-Spin Breaking}
\label{u-spin breaking}

A non-zero strange-quark mass in the strong-interaction Hamiltonian
gives rise to a new U-spin triplet operator that enters the hadronic interactions.
In particular, this can turn $U=0$ operators in the weak effective Hamiltonian into
$U=1$ amplitudes (denoted as $\Delta B_{U=1}^{(')}$)
in the hadronic matrix elements and vice versa ($H_{U=1}$ contributes to $\Delta A_{U=0}$).
The decomposition of the relevant decay amplitudes including (first-order) U-spin breaking then can 
be written as (see also \cite{Pirtskhalava:2011va,Jung:2012ip})
\begin{align}
{\cal A}[D^0 \to K^- \pi^+ ] & \equiv  2 \,V_{cs}^* V_{ud} \left( B_{U=1}  - \Delta B_{U=1}' \right) 
=  2 \,V_{cs}^* V_{ud} \, B_{U=1} \left[  1 - r_1' \, e^{i \, \phi_1'} \right]
\,,
\nonumber \\[0.3em]
{\cal A}[D^0 \to \pi^+ \pi^-] &= (\lambda_d + \lambda_s) \, ( A_{U=0} + \Delta B_{U=1}) +
                                 (\lambda_d -\lambda_s) \, (B_{U=1} + \Delta A_{U=0}) \,
\nonumber \\[0.1em] 
&= B_{U=1}\left[(\lambda_d + \lambda_s)\,\left(r\, e^{i\,\phi} + r_1\,e^{i\,\phi_1} \right) + 
                    (\lambda_d -\lambda_s) \, \left(1 + r_0\, e^{i\,\phi_0}\right)\right]\,, \nonumber 
\\[0.3em] 
{\cal A}[D^0 \to K^+ K^-] &= (\lambda_d + \lambda_s) \, (A_{U=0}- \Delta B_{U=1})
     - (\lambda_d -\lambda_s) \, ( B_{U=1} - \Delta A_{U=0}) \, 
\nonumber \\[0.1em]
&= B_{U=1}\left[(\lambda_d + \lambda_s)\, \left(r\, e^{i\,\phi} - r_1\,e^{i\,\phi_1} \right) 
             - (\lambda_d -\lambda_s) \, \left(1 - r_0\, e^{i\,\phi_0}\right)\right] \,,
\nonumber \\[0.3em]
 {\cal A}[D^0 \to K^+ \pi^-] &= 2 \,V_{cd}^* V_{us} \left(  B_{U=1} + \Delta B_{U=1}' \right)
=  2 \,V_{cd}^* V_{us} \, B_{U=1} \left[  1 + r_1' \, e^{i \, \phi_1'} \right] \,.
\label{amp_breaking}
\end{align}
In the second equations of this parametrization, we have introduced various amplitude ratios and
strong phases with respect to the originally leading $B_{U=1}$ amplitude.

It is important to note that in the SM  --- or for that matter in many NP models ---  direct CP asymmetries 
in $D^0\to K^\mp \pi^\pm$ cannot arise due to the absence of penguin contributions in these decay modes.
However, $D^0$ and $\bar D^0$ both can decay to the same final state $K^\mp \pi^\pm$ which
leads to interference between $D^0$-$\bar D^0$ mixing and the decay processes.
 Such a contribution though is expected to be extremely small (see below).
On the other hand, if there are charged scalars with non-trivial flavour couplings and CP-odd phases
(like, for instance in general models with extended Higgs sector), 
then direct CP asymmetries may arise in $D^0 \to K^\mp \pi^\pm$, too. 
The largish BR of about 3.8\%\ for $D^0 \to K^- \pi^+$ 
thus provides a valuable opportunity to search for a non-standard CP phase.  LHCb
and  Super-B factories producing about $10^9$ charm mesons should have a 5$\sigma$ 
reach of searching for CP asymmetries at the level of about 0.2\% in this mode
(similarly, for $D^0 \to K^+\pi^-$ with a BR of about $1.5 \cdot 10^{-4}$
one can search for $\ord(1\%)$ CP asymmetries).
Note also that this simple decay mode is of crucial importance in extracting the CKM angle $\gamma$ 
from the $B \to D K$ decays in the ADS~\cite{ADS} analysis, where traditionally 
one assumes no CP asymmetry in the subsequent D-decays. 
A confirmation of this assumption by direct experimental searches is therefore very valuable
in any case.

\subsection{Phenomenological Constraints on Amplitude Ratios}

\begin{figure}[t!bpt] 
\begin{center} 
\fbox{\includegraphics[height=0.2\textheight]{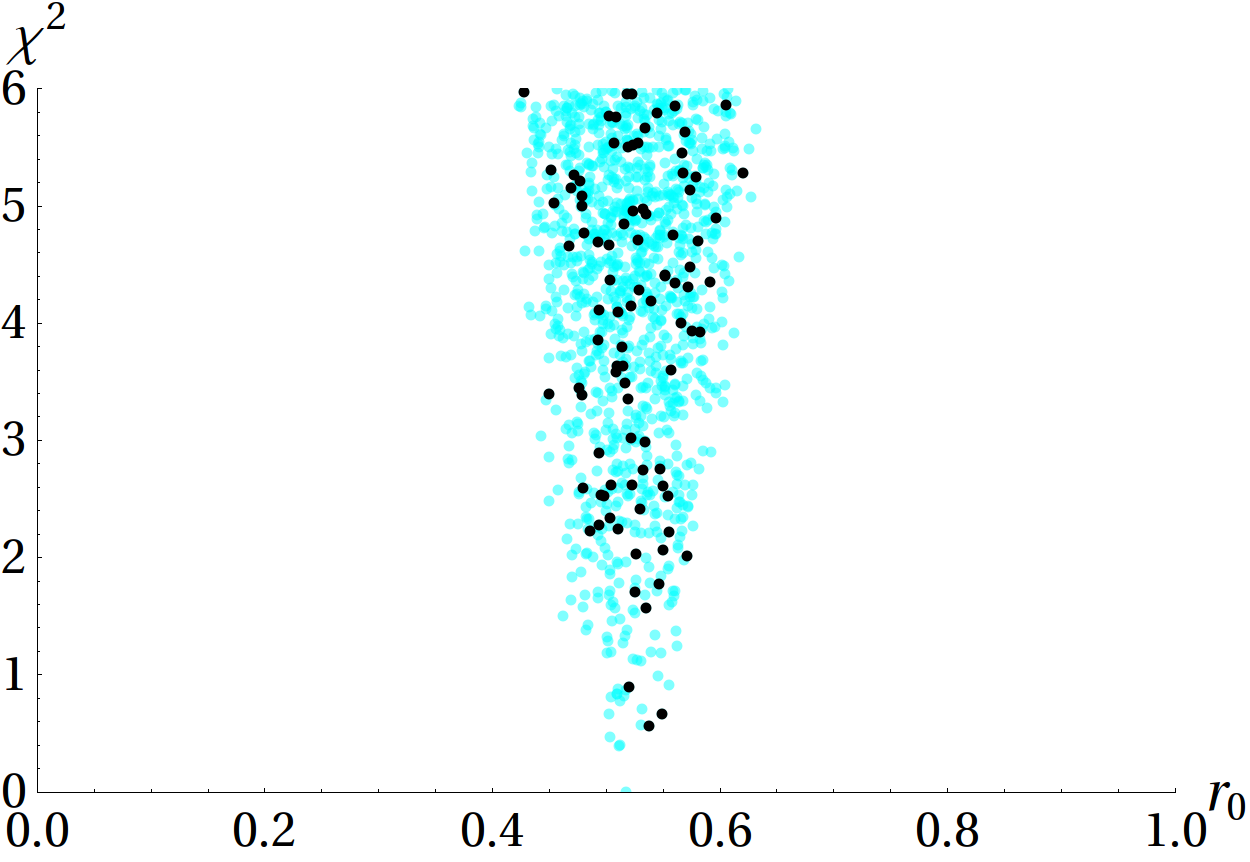}}
\fbox{\includegraphics[height=0.2\textheight]{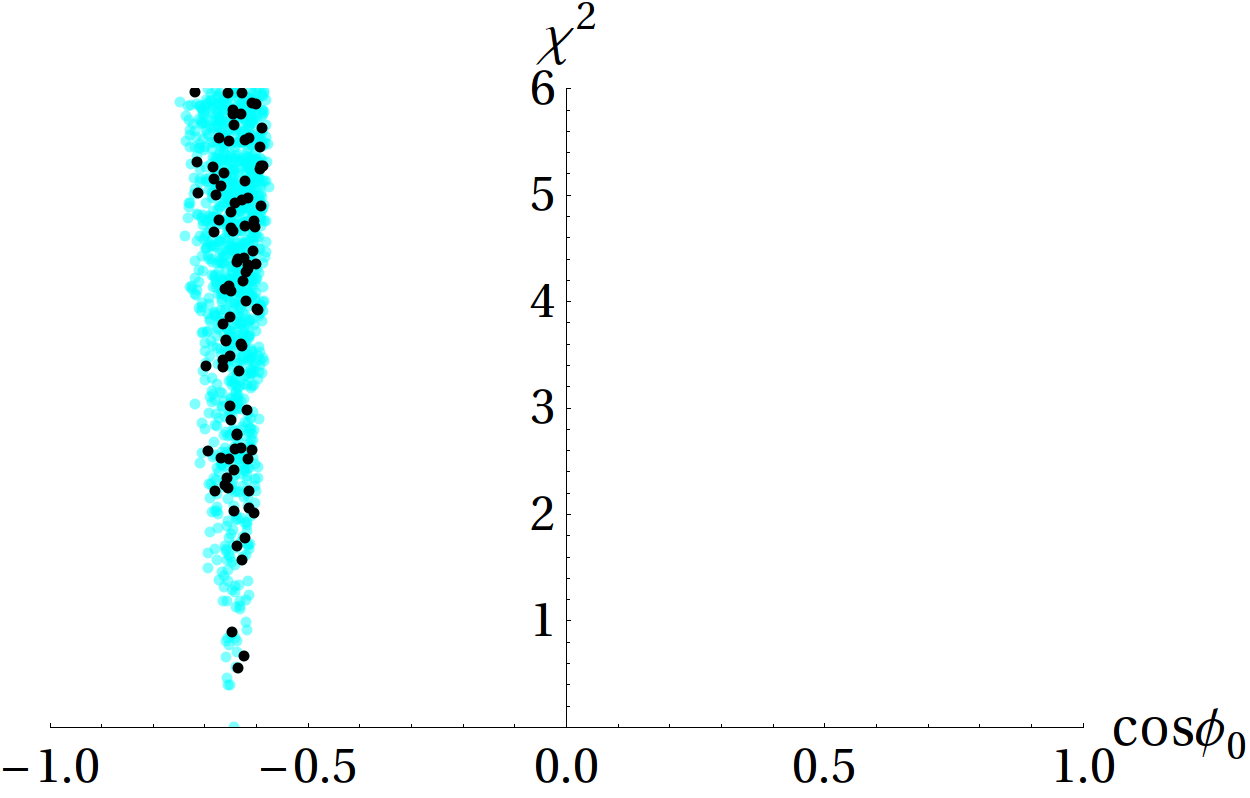}}
\\
\fbox{\includegraphics[height=0.198\textheight]{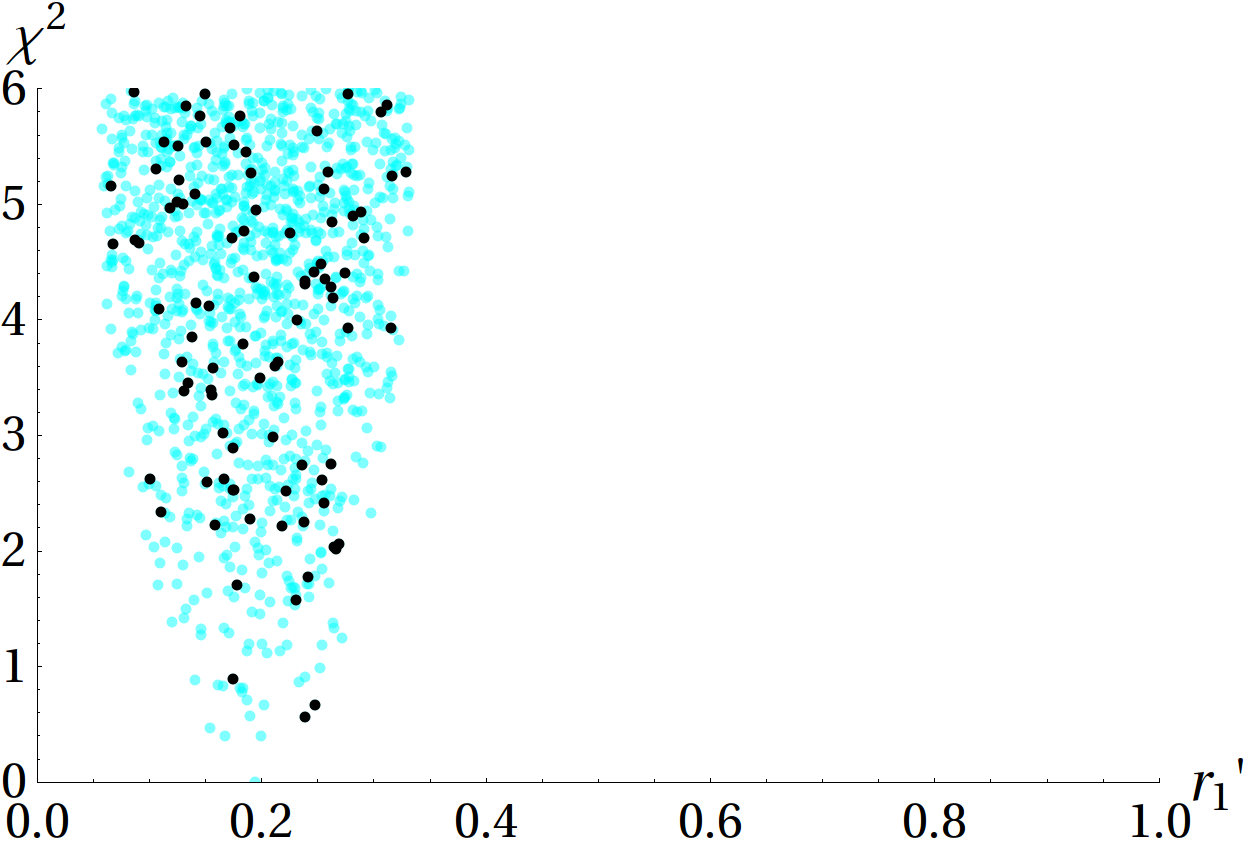}}
\fbox{\includegraphics[height=0.198\textheight]{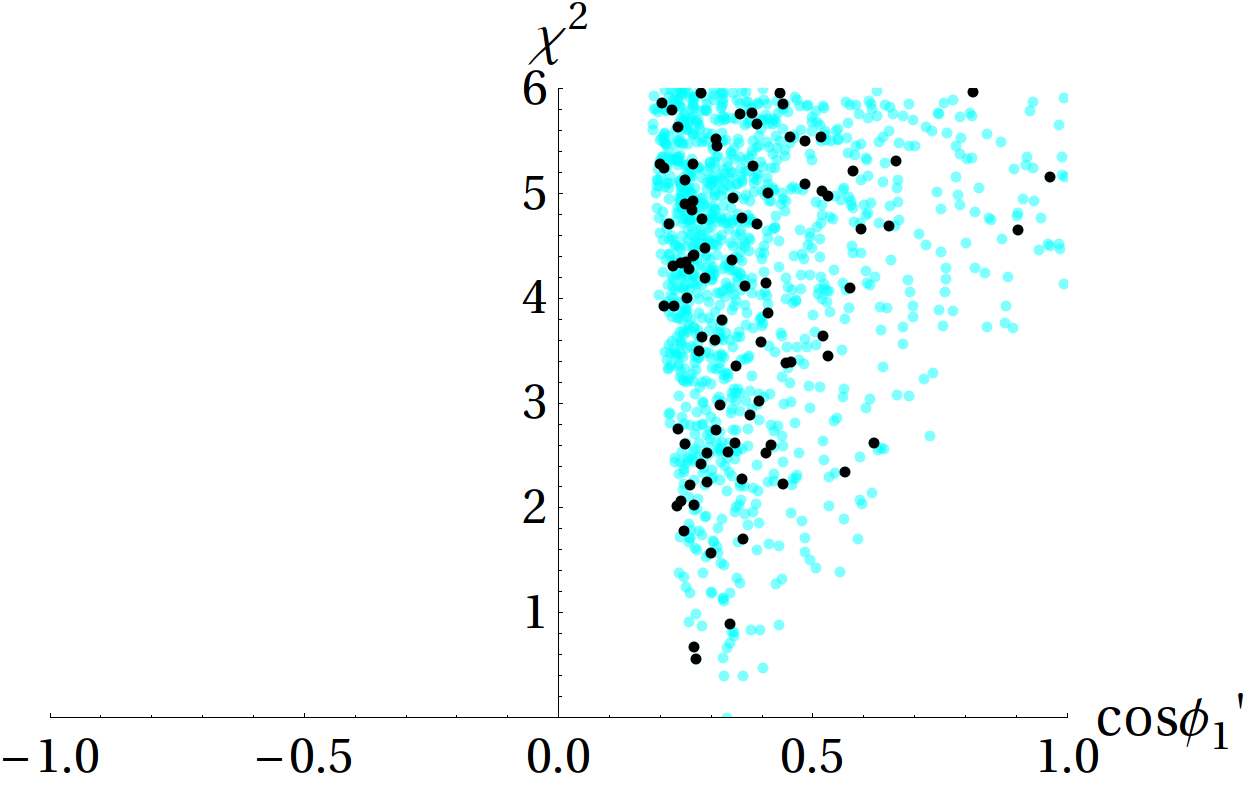}}
\end{center}
 \caption{Fit result for the amplitude parameters $r_0$ and $\cos\phi_0$ (upper row) and
$r_1'$ and $\cos\phi_1'$ (lower row), determining
the amount of U-spin breaking in $D^0 \to P^+ P^-$ BRs.
The generic points shown in light blue are consistent with the experimental constraints
at the $2\sigma$-level and obey $\chi^2\leq 6$.
The black points denote a subset of points where the strong phase differences
between $A_{U=0}$ and $\Delta A_{U=0}$, as well as between $B_{U=1}$ and $\Delta B_{U=1}$
are assumed to be equal within a few percent, 
$(\phi - \phi_0) = \{0,\pi\}$  and $\phi_1=\{0,\pi\}$.}
 \label{fig:Uspin1}
\end{figure}

\begin{figure}[t!bpt] 
\begin{center} 
\fbox{\includegraphics[width=0.48\textwidth]{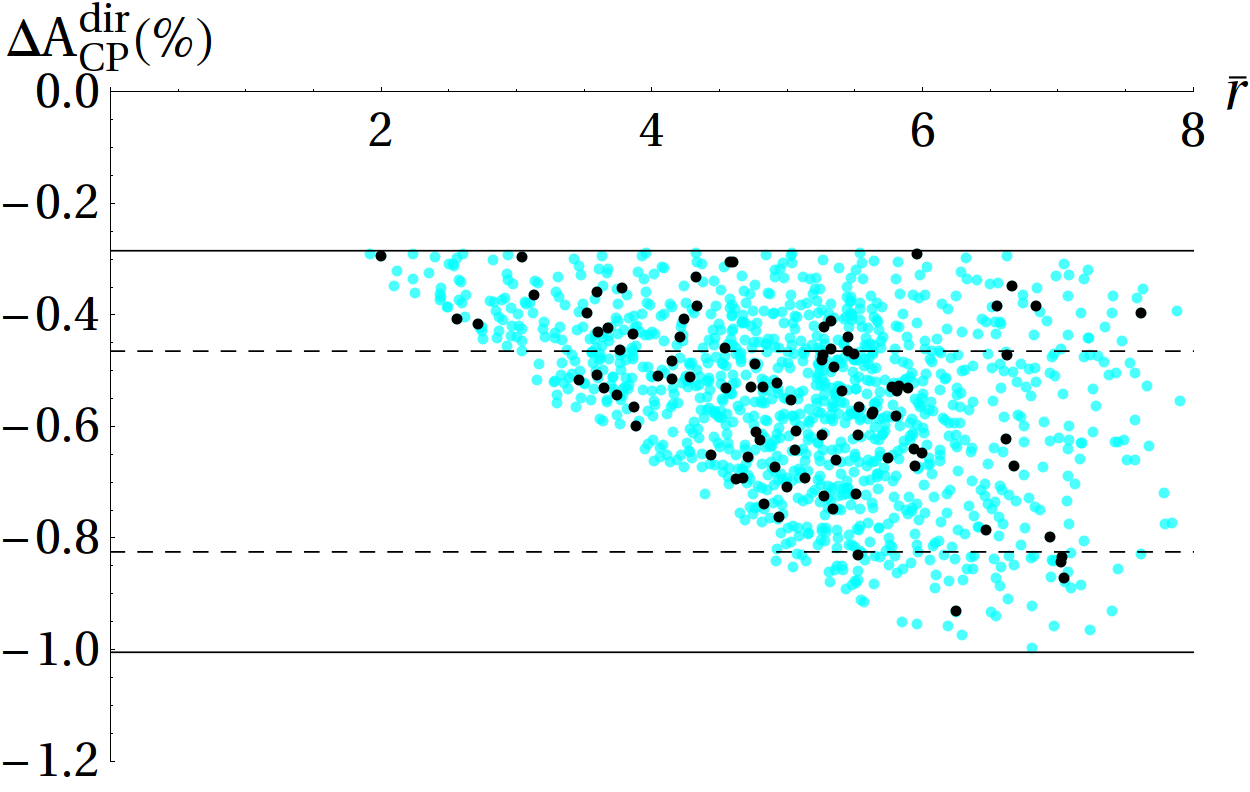}}\\[0.5em]
\fbox{\includegraphics[width=0.48\textwidth]{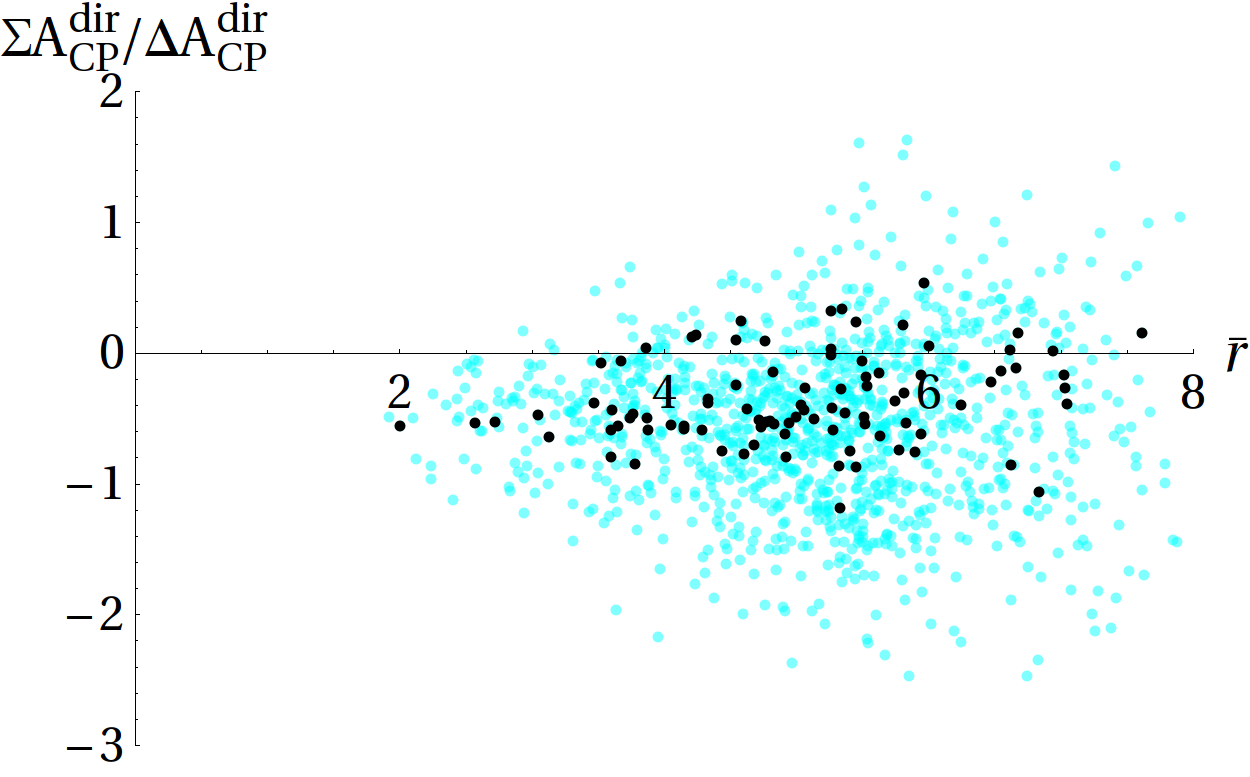}}\\[0.5em]
\fbox{\includegraphics[width=0.48\textwidth]{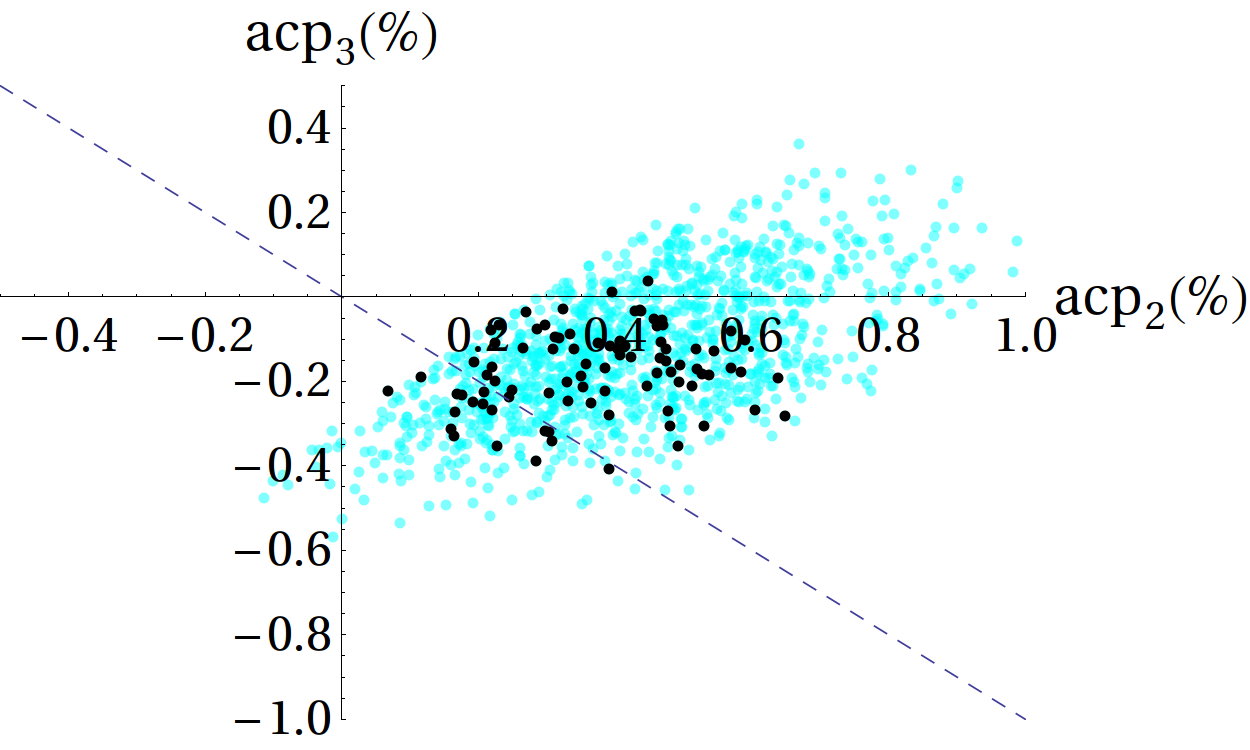}}
\end{center}
 \caption{Fit result for difference of CP asymmetries $\Delta A_{\rm CP}^{\rm dir}$
(top; horizontal lines indicate the $1\sigma$ and $2\sigma$ experimental constraints),
and the ratio of the sum and difference of CP asymmetries (center) in $D^0\to K^+K^-$
and $D^0 \to \pi^+\pi^-$ decays as a function of $\bar r=\sqrt{r^2/2+r_1^2/2}$. 
The lower plot shows the correlation between the two asymmetries ${\rm acp_3}=A_{\rm CP}^{\rm dir}(D^0 \to K^+K^-)$
and ${\rm acp_2}=A_{\rm CP}^{\rm dir}(D^0 \to \pi^+\pi^-)$, (the dashed line indicates the naive U-spin limit). 
Plot conventions as in fig.~\ref{fig:Uspin1}.}
 \label{fig:Uspin2}
\end{figure}

In the following, we perform a fit to the 8 amplitude parameters ($r,r_0,r_1,r_1'$ and $\phi,\phi_0,\phi_1,\phi_1'$)
to the 3 experimentally measured ratios of BRs in (\ref{obs1}--\ref{obs3})
and the difference of CP asymmetries in \eqref{hfag_avg}. 
We also take into account the  fitted value \cite{hfag}
of the strong phase difference between the $D^0\to K^\pm \pi^\mp$ decays,
$\Delta \phi=22.4^\circ{}^{+\phantom{1}9.7^\circ}_{-11.0^\circ}$.
To this end, we generated
random points  which (i) lie within the $2\sigma$ ranges for each experimental observable, and
(ii) yield a total $\chi^2$-value of less than  $6$. We have allowed arbitrary strong phases while the
magnitudes of the various amplitude ratios are considered in a conservative range $|r_X| \leq 8$.
The result is illustrated in figs.~\ref{fig:Uspin1},\ref{fig:Uspin2}.

From fig.~\ref{fig:Uspin1}, we observe that the parameters describing the U-spin breaking effects
from the amplitudes  $\Delta A_0$ and $\Delta B_1'$ are rather tightly constrained, with 
\begin{align*}
  r_0 \simeq 0.52 \,, \quad \cos\phi_0 \simeq -0.64 \,, \qquad r_1' \simeq 0.19 \,, \quad \cos\phi_1' \gtrsim 0.18 \,.
\end{align*}
The individual values for the magnitudes ($r,r_1$) and phases ($\phi,\phi_1$) related to the 
Cabibbo-suppressed operators are not fixed from the fit. 
A useful parameter to study the effect on the direct CP asymmetries is given by
the average
\begin{align}
 \bar r =\sqrt{r^2/2+r_1^2/2} \,.
\end{align}
This is illustrated in fig.~\ref{fig:Uspin2},
where we have plotted the difference of CP asymmetries $\Delta A_{\rm CP}^{\rm dir}$,
as well as their sum divided by the difference as a function of $\bar r$. 
The following interesting observations can be made:
\begin{itemize}
 \item The effect of U-spin breaking from $r_1'$ (entering the BRs for  $D^0 \to K^\pm \pi^\mp$)
is relatively small as compared to the effect from $r_0$ (entering the BRs for $D^0 \to K^+ K^-$ and $D^0 \to \pi^+\pi^-$). 
Again, this can be partly understood in the framework of naive factorization, where in the former case the U-spin breaking in 
the $D \to K(\pi)$ form factors tends to compensate the effect in the decay constants $f_K(\pi)$,
whereas in the latter case, the two effects tend to add up.

 \item There is a clear correlation between $\Delta A_{\rm CP}^{\rm dir}$ and 
     the minimal value for the effective amplitude parameter $\bar r$.
    To reproduce the experimental $2\sigma$-range for $\Delta A_{\rm CP}^{\rm dir}$, one has to require 
   $\bar r \gtrsim 2 $. It remains to be seen whether such large values
    are consistent with our theoretical expectations, given our restricted knowledge on non-perturbative strong dynamics in 
       time-like hadronic processes.
        
 \item The U-spin relation,
      $\Sigma A_{\rm CP}^{\rm dir} = A_{\rm CP}^{\rm dir}(D^0 \to K^+K^-) + A_{\rm CP}^{\rm dir}(D^0 \to \pi^+\pi^-)=0$,
      receives corrections of the same order as $\Delta A_{\rm CP}^{\rm dir}$ itself, or even larger
      (although the case $\Sigma A_{\rm CP}^{\rm dir} =0$ is not excluded).
      Consequently, the correlation between 
       $A_{\rm CP}^{\rm dir}(D^0 \to K^+K^-)$ and $A_{\rm CP}^{\rm dir}(D^0 \to \pi^+\pi^-)$
      can be quite different compared to the naive U-spin limit.
 \item The qualitative results for the BRs and CP asymmetries as a function of the amplitude ratios is well represented
   by a subset of parameters (indicated by the black points in figs.~\ref{fig:Uspin1},\ref{fig:Uspin2})
    where the strong phase differences between the amplitudes $A_{U=0}$ and $\Delta A_{U=0}$
   (as well as between $B_{U=1}$ and $\Delta B_{U=1}$) are set approximately to zero or $\pi$.
\end{itemize}

\subsection{Simplified Analysis of U-Spin Breaking}

As we have seen above, the strong phase differences between
individual $U=1$ (or $U=0$) amplitudes does not play an essential role
for the analysis of direct CP asymmetries in $D^0$ decays (they simply
reflect the redundancy in the effectively 3-dimensional space of the
constrained parameter space in the fit). 
To simplify the further analysis,
we will therefore set 
 \begin{align} \phi_1=(0,\pi) \,, \qquad \phi = (\phi_0,\phi_0 + \pi) \,.
\end{align}

For the description of the BRs, we can also safely neglect the terms proportional to $(\lambda_d+\lambda_s) \sim \ord(\lambda^5)$.
With this approximation, the parameters $r_0$ and $\phi_0$ can be extracted from the observed ratios of BRs, ${\rm obs}_{1-3}$.
The constraint from $\rm obs_1$ alone translates into the inequalities,
\begin{align}
& 0.28 \simeq 1- \frac{2}{\sqrt{{\rm obs}_1}+1} \leq r_0  
\leq 1+ \frac{2}{\sqrt{{\rm obs}_1}-1} \simeq 3.51 \,, 
\nonumber \\[0.25em]
&
 \cos\phi_0  \leq - \frac{{\rm obs}_1-1}{{\rm obs}_1+1} \simeq - 0.53 \,.
\end{align}
Notice that these bounds already constrain the possible values for $\rm obs_2$ and $\rm obs_3$.
Including the average of  $\rm obs_2$ and $\rm obs_3$, the central
values of $r_0$ and $\phi_0$ take values
\begin{align}
r_0 & \simeq 0.54  \,,
\qquad  
\cos\phi_0  \simeq -0.69 \,,
\label{leading}
\end{align}
which are consistent with the previous fit within the uncertainties.\footnote{
The result of this procedure is  similar to a recent analysis of $SU(3)$ flavour-symmetry
breaking in non-leptonic $D$ decays considered in \cite{Pirtskhalava:2011va}, 
where the authors neglected  what would correspond to our parameter $r_1'$ 
which has been motivated by dropping higher $SU(3)$ representations in the
effective Hamiltonian.}

Concerning the direct CP asymmetries, the constraints from ${\rm obs}_{1-3}$ alone cannot
restrict the Cabibbo-suppressed amplitudes $A_{U=0}$ or $\Delta B_{U=1}$ which only provide
a tiny ($<1\%$) correction to the BRs. Within our approximations, and assuming that all
amplitude ratios are of ${\cal O}(1)$, we obtain
\begin{align}
A_{\rm CP}^{\rm dir}(K^+K^-)- A_{\rm CP}^{\rm dir}(\pi^+\pi^-) & \simeq 
\frac{8 \, {\rm Im}[\lambda_d \lambda_s^*]}{|\lambda_d-\lambda_s|^2} \cdot
\frac{ \left(1+r_0^2\right)\left( -r \sin\phi + r_0 r_1 \sin(\phi_0-\phi_1) \right)}{1 - 2 r_0^2 \cos 2\phi_0 + r_0^4} 
\,,
\label{acpdiff}
\end{align}
together with 
\begin{align}
  \frac{A_{\rm CP}^{\rm dir}[D^0 \to K^+K^-]+A_{\rm CP}^{\rm dir}[D^0 \to \pi^+\pi^-]}{A_{\rm CP}^{\rm dir}[D^0 \to K^+K^-]-A_{\rm CP}^{\rm dir}[D^0 \to \pi^+\pi^-]}
 & \simeq - \frac{{\rm obs}_1-1}{{\rm obs}_1+1} \simeq -53\% \,.
 \label{break}
 \end{align}
The latter equation shows again that --- as soon as U-spin violating effects are included --- the
direct CP asymmetries in $D^0\to \pi^+\pi^⁻$ and $D^0 \to K^+K^-$ are no longer equal in magnitude.
The particular value in (\ref{break}) reproduces the central value of the trend apparent in
fig.~\ref{fig:Uspin2} (right); allowing for arbitrary strong phases, the effect can become even larger.
The former eq.~\eqref{acpdiff} states that the measured difference of direct CP asymmetries (considering
the leading term in the Wolfenstein expansion) is a product of 3 terms,
\begin{align}
 f_{\rm weak} &= 4 \, {\rm Im}\left[\frac{\lambda_d^*+\lambda_s^*}{\lambda_d-\lambda_s} \right] 
\simeq  
2 A^2 \lambda^4 \eta 
 \approx  0.11 \% \,,
\nonumber \\[0.25em]
 f_{\rm \Delta U} &= \frac{ 1+r_0^2}{1 - 2 r_0^2 \cos 2\phi_0 + r_0^4} \simeq 1.1
\nonumber \\[0.25em]
 f_{\rm strong} &= -r \sin\phi + r_0 r_1 \sin\phi_0 \cos\phi_1 \,.
\end{align}
In order to achieve the central value of the experimentally observed number, $f_{\rm strong}$ should be about $-5.3$,
which requires accordingly large values for $r$ and $r_1$. 
More moderate values for $r,r_1$ are possible if  we consider the 2$\sigma$ range, only (as we have seen before).
As the direct CP asymmetries scale linearly with both the amplitude ratios $r$ and $r_1$, there is no
preference for whether the original amplitude $A_{U=0}$ or the effect of U-spin violation through
$\Delta B_{U=1}$ should be dominating the required value of $f_{\rm strong}$, although the
effect of $r_1$ is somewhat diminished by the factor $r_0 <1$.

\begin{figure}[t!bpt] 
\begin{center} 
\fbox{\includegraphics[width=0.45\textwidth]{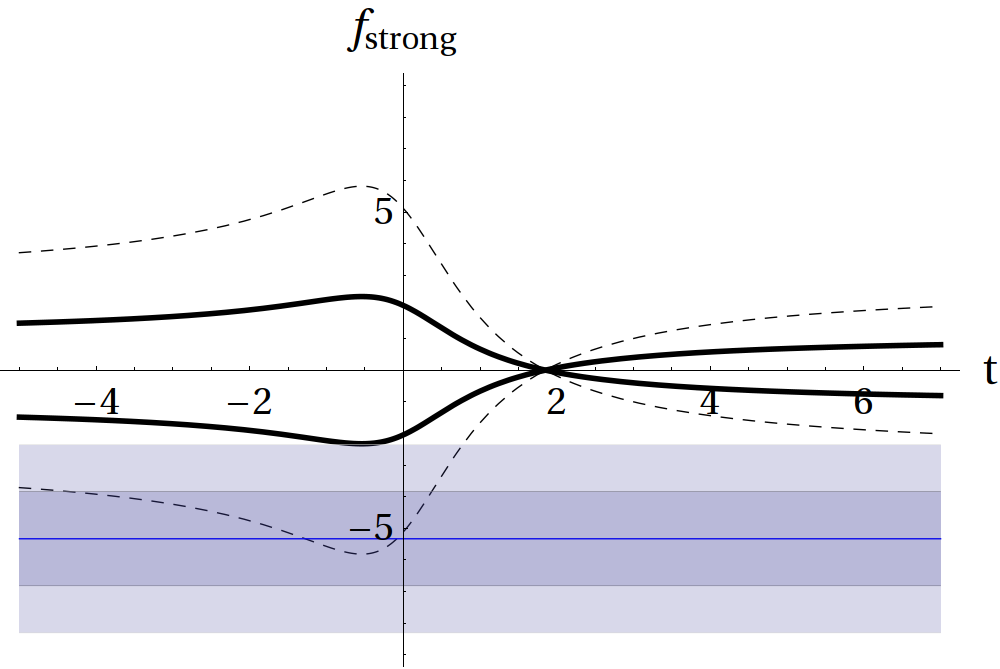}}
\end{center}
 \caption{The function $f_{\rm strong}$, determining the size of $\Delta A_{\rm CP}^{dir}$ in the SM,
  as a function of the amplitude ratio $t$ for two different values of the averaged amplitude ratios,
   $\bar r =2$ (solid lines) and $\bar r=5$ (dashed lines), 
  see eq.~\eqref{effpar}. The values necessary to recover the experimental 1$\sigma$ (2$\sigma$) range 
  in eq.~\eqref{hfag_avg} are shown in different shades of grey.
 \label{fig:plotSM}}
\end{figure}
To illustrate the result, we consider the average of amplitude parameters $\bar r$ together with 
\begin{align}
 t \equiv \frac{r_1 \,  \sin\phi_0\cos \phi_1 }{r \, \sin\phi} 
 = \pm \frac{r_1}{r} \,.
\label{effpar}
\end{align}
In fig.~\ref{fig:plotSM}, we plot the result for $f_{\rm strong}$ together with the experimentally
preferred range for two values $\bar r =\{2,5\}$ as a function of $t$.
As before, we can see that with $\bar r=5$ (dashed line) and for negative values of $t$ the result
falls comfortably into the experimental $1\sigma$ range, whereas for $\bar r=2$ (solid line), the
result is marginally consistent with the measurement at the $2\sigma$ level.

\subsection{A Note on Mixing-induced CP Asymmetries}

As described in eq.~\eqref{acpf}, apart from the direct CP asymmetry, the time-integrated CP asymmetry $A_{\rm CP}(f)$ 
in the decay of $D^0$ to a given final state $f$ also contains a part which is due to the interference of 
decay with mixing, the so-called mixing-induced CP asymmetry $A_{\rm CP}^{\rm ind}$ which is defined 
as   
\begin{equation}
A_{\rm CP}^{ind} = - \frac{2\,x\,|R_f|\,(1 + |R_f|^2\,|R_m|^2 )}{(1+|R_f|^2)^2\,|R_m|} \, \sin\phi \,.
\end{equation}
Here $R_f = \frac{{\bar A}_f}{A_f}$ denotes the ratio of the amplitudes for a given decay and its CP conjugate, 
$R_m = \frac{q}{p}$ gives the ratio of the parameters $p,q$ in $D^0$-$\bar D^0$ mixing, 
$x = \frac{\Delta M_D}{\Gamma_D}$ denotes the ratio of the mass difference and decay width in
the  $D^0$-$\bar D^0$ system, 
and  the phase $\phi$ denotes the argument of the function $\lambda_f = \frac{q}{p} \, \frac{{\bar A}_f}{A_f}$.
The present experimental bound on the $D^0-{\bar D}^0$ mixing parameters, i.e.\ the value of $x$ and the phase and magnitude 
of $R_m$ are given by \cite{hfag},
\begin{equation}
x = (0.98^{+0.24}_{-0.26})\,\%\,, \qquad \arg\left(\frac{q}{p}\right) = (- 8.5 ^{+7.4}_{-7.0})^{\circ}\,,\qquad |R_m|=0.87^{+0.17}_{-0.15}\,.
\label{mixexp}
\end{equation}

For decay modes like $D^0\to K^+K^-,\pi^+\pi^-$, which --- in the SM ---- do not have interfering amplitudes with different weak and strong phases 
of comparable size, one has $|R_f| \approx 1$ and $\arg(R_f)\approx 0$.
Hence, the phase $\phi$ will be purely induced by $D^0$-$\bar D^0$ mixing. 
In such cases $A_{\rm CP}^{\rm ind}$ can be treated as a universal quantity which 
does not depend on the final state mesons. In particular, its effect will drop
out in the difference of CP asymmetries, $\Delta A_{\rm CP}$, apart from
a term correcting for the finite experimental cut on the proper decay time, relative
to the life-time associated to the individual decay modes (see e.g.\ \cite{talk:Grossman}).
We have checked that the presence of U-spin violating effects in $A_{\rm CP}^{\rm ind}(K^+K^-)$
and $A_{\rm CP}^{\rm ind}(\pi^+\pi^-)$
have a negligible numerical effect of the order $\lsim 10^{-5}$ on $\Delta A_{\rm CP}$.

\subsection{$\Delta A_{\rm CP}^{\rm dir}$ in the Presence of a Fourth Generation}

In the 4G extension of the SM,
the presence of the heavy $b'$-quark gives an extra contribution to the penguin operators, and it
also modifies the CKM elements, such that $\lambda_d +\lambda_s +\lambda_b = - \lambda_{b'} \neq 0$.
Both effects can be accounted for by including an additional $U=0$ operator 
proportional $\lambda_{b'}$ in the effective Hamiltonian.
The amplitudes in eq.~\eqref{amp_breaking} can then be generalized to the 4G case
as follows, 
\begin{align}
{\cal A}[D^0 \to \pi^+ \pi^-] &= B_{U=1}\left[
(-\lambda_b) \left(r\, e^{i\,\phi} + r_1 \, e^{i\phi_1}\right) +
(\lambda_d -\lambda_s)  \left(1 + r_0\, e^{i\,\phi_0}\right) \right. 
\nonumber \\[0.1em] & \left. - 
\lambda_{b'} \left( r_4\,e^{i\,\phi_4} +  r_4' \, e^{i \, \phi_4'} \right) 
\right]\,, 
\nonumber \\[0.3em] 
{\cal A}[D^0 \to K^+ K^-] &= B_{U=1}\left[ 
(-\lambda_b) \left(r\, e^{i\,\phi} - r_1 \, e^{i\phi_1}\right) - 
(\lambda_d -\lambda_s) \, \left(1 - r_0\, e^{i\,\phi_0}\right) \right.
\nonumber \\[0.1em] & \left. - 
\lambda_{b'} \left( r_4\,e^{i\,\phi_4} -  r_4' \, e^{i \, \phi_4'} \right)
\right],
\label{amp_breaking4}
\end{align}
where $r_4^{(')}$ and $\phi_4^{(')}$ parametrize the contribution of the new
$U=0$ operator and its $U=1$ counterpart from U-spin violation in the corresponding
hadronic matrix elements. As before, we assume that the corresponding strong phases
are related to the other $U=0$ and $U=1$ amplitudes.
Then, the additional contribution to $\Delta A_{\rm CP}^{\rm dir}$ can again be factorized,
\begin{align}
\Delta A_{\rm CP}^{\rm dir} & \simeq f_{\Delta U} \cdot 
 \left\{  f_{\rm weak} \cdot f_{\rm strong} + f_{\rm weak}^{\rm 4G}  \cdot f_{\rm strong}^{\rm 4G} \right\} 
\nonumber\\[0.1em]
 & \equiv  f_{\Delta U} \cdot f_{\rm weak} \cdot f_{\rm strong}^{\rm eff} \,,
\end{align}
with 
\begin{align}
 f_{\rm weak}^{\rm 4G} &=  4 \, {\rm Im}\left[\frac{\lambda_{b'}}{\lambda_d-\lambda_s} \right]  
 \simeq   \frac{2 \, \sin\theta_{14} \sin\theta_{24} \sin(\delta_{14} - \delta_{24}) }{\sin\theta_{12}} \,,
\nonumber \\[0.25em]
 f_{\rm strong}^{\rm 4G} &= -r_4 \sin\phi_4 + r_0 r_4' \sin\phi_0\cos\phi_4'  \,.
\end{align}
Here, we have used the PDG-type parametrization for the 4G CKM matrix as in
\cite{Buras:2010pi}, and neglected higher-order terms in the Wolfenstein expansion.
Notice that eq.~\eqref{break} is still valid if our assumptions on the strong phases hold.

\begin{figure}[t!] 
\begin{center} 
\fbox{\includegraphics[width=0.45\textwidth]{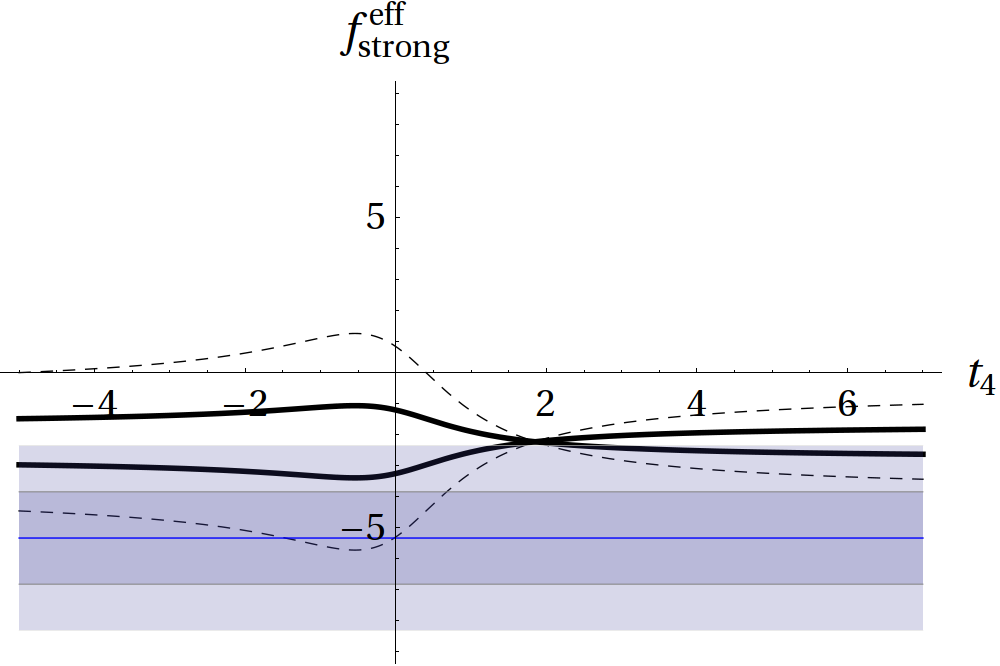}}
\fbox{\includegraphics[width=0.45\textwidth]{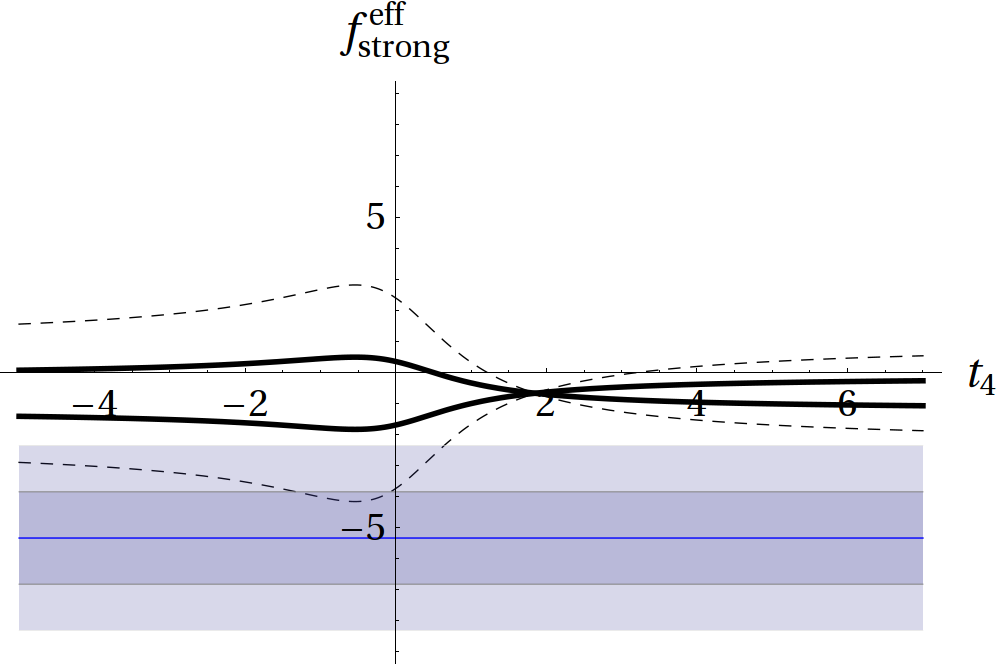}}
\end{center}
\caption{The function $f_{\rm strong}^{\rm eff}$, determining the size of $\Delta A_{\rm CP}^{dir}$ in the 4G extension
  of the SM,
  as a function of the amplitude ratio $t_4$ for two different values of the averaged amplitude ratios,
   $\bar r_4 =1$ (solid lines) and $\bar r_4=3$ (dashed lines). The associated weak factor is set equal
  to the SM one, $f_{\rm weak}^{\rm 4G}= f_{\rm weak}$, and the SM amplitude ratios are set
  to $r=r_1=2$. The left (right) plot refers to a choice of strong phases where the two SM contributions $r$ and $r_1$
  combine constructively (destructively) to a value of $-2.2$ ($-0.67$).
  The values necessary to recover the experimental $1\sigma$ and $2\sigma$ range in eq.~\eqref{hfag_avg}
  are shown in different shades of grey.
\label{fig:SM4}}
\end{figure}

From the analysis of various flavour observables in the kaon and $B$-meson sector 
(see e.g.\ \cite{Soni:2008bc,Bobrowski:2009ng,Soni:2010xh,Buras:2010pi}), we know 
that the product of mixing angles $\theta_{14}$ and $\theta_{24}$ can be as large as $\ord(\lambda^4)$, up-to $\ord(\lambda^3)$,
and therefore, in principle, can lead to an enhancement of the CP asymmetries by up to one order of magnitude
as compared to the SM (where $\sin\theta_{13}\sin\theta_{23} \sim \ord(\lambda^5)$).
However, it has also been found that for such large values of 4G mixing angles, the new CP phases $\delta_{i4}$
are rather fine-tuned to values satisfying $\delta_{14}\simeq \delta_{24}$
\cite{SNAS,Buras:2010pi}. 
This implies that the 4G CKM elements \emph{cannot} lead to a substantial \emph{parametric} enhancement
(i.e.\ a suppression factor with less powers of the Wolfenstein parameter $\lambda$) of
CP asymmetries in $D$ decays (contrary to a recent claim in \cite{Rozanov:2011gj}, where --- to our understanding ---
the phenomenological constraints 
on the 4G CP phases have not been taken into account properly). 
Still, the presence of the additional penguin operators can lead to 
a numerical enhancement which, however, is hard to quantify reliably 
without more detailed knowledge on the 
hadronic amplitude parameters.
To illustrate the result, we define again
\begin{align}
 \bar r_4 \equiv \sqrt{\frac{r_4^2+(r_4')^2}{2}} \,, \qquad
 t_4 \equiv \frac{r_4' \,  \sin\phi_0\cos\phi_4' }{r_4 \, \sin\phi_4} \,,
\end{align}
and show in fig.~\ref{fig:SM4} the result for $f_{\rm strong}^{\rm eff}$
as a function of $t_4$ for  $\bar r_4 =\{1,3\}$, while we fix $f_{4G}=f_{\rm weak}$ and $r=r_1=2$,
for simplicity,
and compare two different choices for the associated phases corresponding to constructive/destructive
effects in the SM contribution (assuming an overall negative sign).
One sees that $\Delta A_{\rm CP}^{\rm dir}$ can receive some deviations from the SM with
3 generations, but without a quantitative theoretical estimate of the strong amplitudes,
nothing more can be said.


\subsection{Generic NP Explanation of $\Delta A_{\rm CP}^{\rm dir}$}

The above results can easily be generalized to generic NP models, and
the following statements can be made:
\begin{itemize}
 \item Both,  additional NP operators in the weak Hamiltonian with $U=0$ or $U=1$, can 
        contribute to the direct CP asymmetries.
 \item  The new flavour coefficients can potentially enhance the 
         corresponding factor $f_{\rm weak}^{\rm NP}$ as compared to the SM. 
        As a consequence, the observed value for $\Delta A_{\rm CP}^{\rm dir}$ can be obtained
        with somewhat smaller values for the various hadronic amplitudes.
 \item Even in the presence of NP, in order to simultaneously explain the
  ratios of $D^0$ decay BRs, while keeping the direct CP asymmetries sufficiently
  small, we have to require large U-spin violating effects. Therefore, in general, 
    the direct CP asymmetries in $\pi^+\pi^-$ and $K^+K^-$ 
    will not be equal in magnitude.
\end{itemize}
For more detailed NP analyses, see e.g.\ \cite{Giudice:2012qq,Altmannshofer:2012ur}.

\subsection{Possibilities for Direct CP Searches in Charged Modes}

Given that the current focus is on direct CP violation in neutral $D^0$ 
decay modes, we note in passing that decays of charged  meson, $D^+$ and $D_s$,
also offer many interesting and experimentally distinctive channels for such CP studies,
with the advantage that $D^0$-$\bar D^0$ mixing is not involved.
In table~\ref{tab1}, we list a few  2-body decay modes of $D^+$ and $D_s$ mesons where
a non-vanishing CP asymmetry has been reported with $\gsim 2 \sigma$ significance.
With the foreseen statistics at LHCb and future Super-B factories, we find that
interesting sensitivity to CP-violating effects in the SM or beyond, to the level of
a few per mille,  can be
anticipated for these examples, too.

The current measurement of the CP asymmetry in the CA decay $D^+ \to K_S \pi^+$ 
--- within the present experimental uncertainties --- 
is consistent with the SM expectation from $K^0$-$\bar K^0$ mixing (which is of order 
$\epsilon_K \sim \lambda^3$), while the direct CP asymmetry 
is suppressed by a factor of $\lambda^6$ and therefore negligible
\cite{Bigi:1994aw,Lipkin:1999qz}.
NP contributions to the direct CP asymmetries of the order of several per mille 
could thus alter the SM prediction significantly which could be tested in the future.
In this context, it would also be useful to tag on the flavour eigenstate in
$D^+ \to \bar K^0 \pi^+$ (or also $D^+ \to \bar K^{*0} \pi^+$).
In a similar fashion, in the SCS decay mode $D_s \to K_S \pi^+$, 
a contribution from $K^0$-$\bar K^0$ mixing is unavoidable, but in this case also
penguin effects can contribute to $A_{\rm CP}^{\rm dir}$ at the
per-mille level, as shown in the preceding discussion. The current experimental central value is rather 
high, but only with a small significance of about $2\sigma$. 
Here too, the decay into the flavour state, $D_s \to K^{(*)0}\pi^+$
deserves experimental attention.
Finally, the CA decay $D_s \to \eta'\pi^+$ is a pure tree with vanishing
CP asymmetry in the SM.
Evidently, if the current $2\sigma$ hint for the latter decay mode gets verified, 
close to its current central value, that could be an important  sign of NP.

\begin{table}[t!] 
\begin{center}
\begin{tabular}{|c|c|c|c|}
\hline
Mode  &   BR &   $A_{\rm CP}$ in \% &  $5\sigma$ Reach  \\                                                               
\hline \hline 
$D^+\to K_S \pi^+ $ &  $1.47\times 10^{-2}$ & $-0.52 \pm 0.14$ \cite{hfag} &    $1\times10^{-3}$ 
\\
\hline
$D_s \to \eta' \pi^{+} $ &  $3.94 \times 10^{-2}$ &  $-6.1 \pm 3.0$ \cite{pdg} &  $0.7 \times 10^{-3}$ \\
 & & $-5.5 \pm 3.7 \pm 1.2$  \cite{hfag}  & 
\\
\hline
$D_s \to K_S \pi^+$ &  $1.21 \times 10^{-3} $ &  $6.6 \pm 3.3$ \cite{pdg} &   $4 \times 10^{-3}$ \\
 & &  $ 6.53 \pm 2.46$ \cite{hfag} &  
\\
\hline
\end{tabular}
\caption{BRs and  CP asymmetries in different charged $D^+$ and $D_s$ decay modes. 
The quoted (naive) $5\sigma$ reach for the sensitivity on $A_{\rm CP}$ 
refers to $10^9$ produced $D^+,D_s$ mesons at LHCb or future Super-B factories.}
\label{tab1}
\end{center}
\end{table}

There are many other modes of D and $D_s$ which could be suitable 
for direct CP studies. Some examples of SCS modes involving penguins 
are: $D^+ \to K^+ \bar K^{*0},  K^{*+} \bar K^0$; $D^+ \to \phi \pi^+, \rho^0 \pi^+ , \pi^+ \pi^0 (\eta')$;  
$D_s \to K^+ \phi (\eta'), K^0 (K^{*0}) \pi^+$ and many more.  There are also further CA modes (with no penguin contributions) 
such  as  $D^+ \to \bar K^0 (\bar K^{*0}) \pi^+ $, $D_s \to \phi \pi^+ (K^+)$ etc.  As mentioned previously, 
the SCS modes with penguins and with typical BRs of a few per mille should  
allow definitive CP searches of ${\cal O}(0.5\%)$, and for CA modes w/o penguins we could get to about 0.2\%.


\section{Insights from Non-Leptonic $B$-Meson Decays}

In this section, we are looking for a correspondence between the considered $D^0 \to P^+P^-$ modes
and suitable non-leptonic $B$-meson decays. Obviously, exchanging the roles of up- and down-type quarks
would lead us to consider a symmetry between $u$ and $c$, in the following referred to as $W$-spin.
Of course, we expect such a symmetry to be badly broken, and rather than use the symmetry relations
themselves --- for the sake of this paper --- we would like to study the
 amplitude ratios in a \emph{parametrization} based on $W$-spin
(including its violation to first order), analogous to eq.~\eqref{amp_breaking}.
In particular, we would like to quantify the phenomenologically allowed values for the
various amplitude ratios in such a case, which might give us some guide-line about what
to expect for the corresponding $D^0\to P^+P^-$ case.

With this in mind, we can repeat our previous analysis for the decays
$$
  \bar B_d^0 \to D_s^- \pi^+, D^+ K^-, K^- \pi^+, D_s^- D^+
$$ 
in a straight-forward manner.\footnote{A similar analysis could be performed for 
U-spin related modes,  where the $B_d$-meson in the initial state is replaced 
by $B_s$, and/or the effective Hamiltonian for $b \to s q\bar q'$ is replaced by
the one for $b\to d q\bar q'$.}  
The relevant effective Hamiltonian, describing the $b \to s q \bar q'$ transitions,
can be decomposed as
\begin{align}
 H_{\rm eff}(b \to s c\bar u) &=  -\left(V_{cb} V_{us}^*\right)  H_{W=1}^{(W_3=-1)} \,, 
\\[0.25em]
 H_{\rm eff}(b \to s q\bar q) &=  \left( \frac{V_{ub} V_{us}^* - V_{cb} V_{cs}^*}{\sqrt 2} \right)  H_{W=1}^{(W_3=0)} 
                                + \left( V_{ub} V_{us} + V_{cb} V_{cs} \right)  H_{W=0} \,,
\\[0.25em] 
 H_{\rm eff}(b \to s u\bar c) &=  \left( V_{ub} V_{cs}^* \right) H_{W=1}^{(W_3=+1)} \,. 
\end{align}
Notice that, in this case, the CKM elements scale as
\begin{align}
&  \lambda_c \equiv V_{cb} V_{cs}^* \simeq A \lambda^2 \,, &&
  V_{cb} V_{us}^* \simeq A \lambda^3 \,, 
\nonumber \\[0.3em]
&  \lambda_u \equiv V_{ub} V_{us}^* \simeq A \lambda^4 (\rho - i \eta)
\,, && V_{ub} V_{cs}^* \simeq A \lambda^3 (\rho - i \eta) \,,
\end{align}
which changes the relative importance of the corresponding hadronic amplitude parameters
for BRs and CP asymmetries in comparison with the U-spin analysis of $D^0$ decays.
The latter can be parametrized analogously to eq.~\eqref{amp_breaking},
\begin{align}
{\cal A}[\bar B^0 \to D^+ K^-] &
= 2 \,V_{cb} V_{us}^* \, B_{W=1}
\left[  1 - r_1' \, e^{i \, \phi_1'} \right] \,,
\nonumber\\[0.25em]
{\cal A}[\bar B^0 \to K^- \pi^+] 
&= B_{W=1}\left[(\lambda_u + \lambda_c)\,\left(r\, e^{i\,\phi} + r_1\,e^{i\,\phi_1} \right) + 
                    (\lambda_u -\lambda_c) \, \left(1 + r_0\, e^{i\,\phi_0}\right)\right]\,, 
\nonumber \\[0.25em] 
{\cal A}[\bar B^0 \to D_s^- D^+] 
&= B_{W=1}\left[(\lambda_u + \lambda_c)\, \left(r\, e^{i\,\phi} - r_1\,e^{i\,\phi_1} \right) 
             - (\lambda_u -\lambda_c) \, \left(1 - r_0\, e^{i\,\phi_0}\right)\right] \,,
\nonumber\\[0.25em]
 {\cal A}[\bar B^0 \to D_s^- \pi^+] & 
= 2 \,V_{ub} V_{cs}^*\, B_{W=1}
\left[  1 + r_1' \, e^{i \, \phi_1'} \right]
\,,
\label{amp_breaking2}
\end{align}
where for simplicity we have used the same notation for the amplitude ratios as
for the U-spin analysis of $D^0$ decays.
As for the experimental input, we consider the four observables \cite{pdg}
\begin{align}
{\rm br_1/br_2} & = \frac{{\rm Br}[\bar B^0 \to D^+ K^-]/|\vec p_{DK}|}{{\rm Br}[\bar B^0\to K^-\pi^+]/|\vec p_{K\pi}|} \simeq 11.8 \pm 3.6\,,
\nonumber \\[0.25em]
{\rm br_3/br_2} & = \frac{{\rm Br}[\bar B^0 \to D_s^- D^+]/|\vec p_{D_sD}|}{{\rm Br}[\bar B^0\to K^-\pi^+]/|\vec p_{K\pi}|} \simeq 536 \pm 62\,,
\nonumber \\[0.25em]
{\rm br_4/br_2}& =\frac{{\rm Br}[\bar B^0 \to D_s^- \pi^+]/|\vec p_{D_s\pi}|}{{\rm Br}[\bar B^0\to K^-\pi^+]/|\vec p_{K\pi}|} \simeq 1.28 \pm 0.16 \,,
\end{align}
and
\begin{align}
{\rm acp_2} = A_{\rm CP}^{\rm dir}[\bar B^0\to K^-\pi^+] & \simeq -0.098 \pm 0.013\,.
\end{align}
From this we can perform a fit to the 8 real parameters ($r,r_0,r_1,r_1'$ and $\phi,\phi_0,\phi_1,\phi_1'$),
where again we generate random points which fulfill the experimental constraints within at least $2\sigma$ and
lead to a total $\chi^2$-value less than 6. To reproduce the difference between the two tree decays
$\bar B^0 \to D^+K^-$ and $\bar B^0 \to D_s^-\pi^+$, in principle, one can allow for two different classes
of solutions, with either $r_1'>1$ or $r_1'<1$.
In the following, we will focus on the case $r_1' <1$, which would be the
natural choice in the factorization approximation.

\begin{figure}[t]
 \begin{center}
  \fbox{\includegraphics[width=0.48\textwidth]{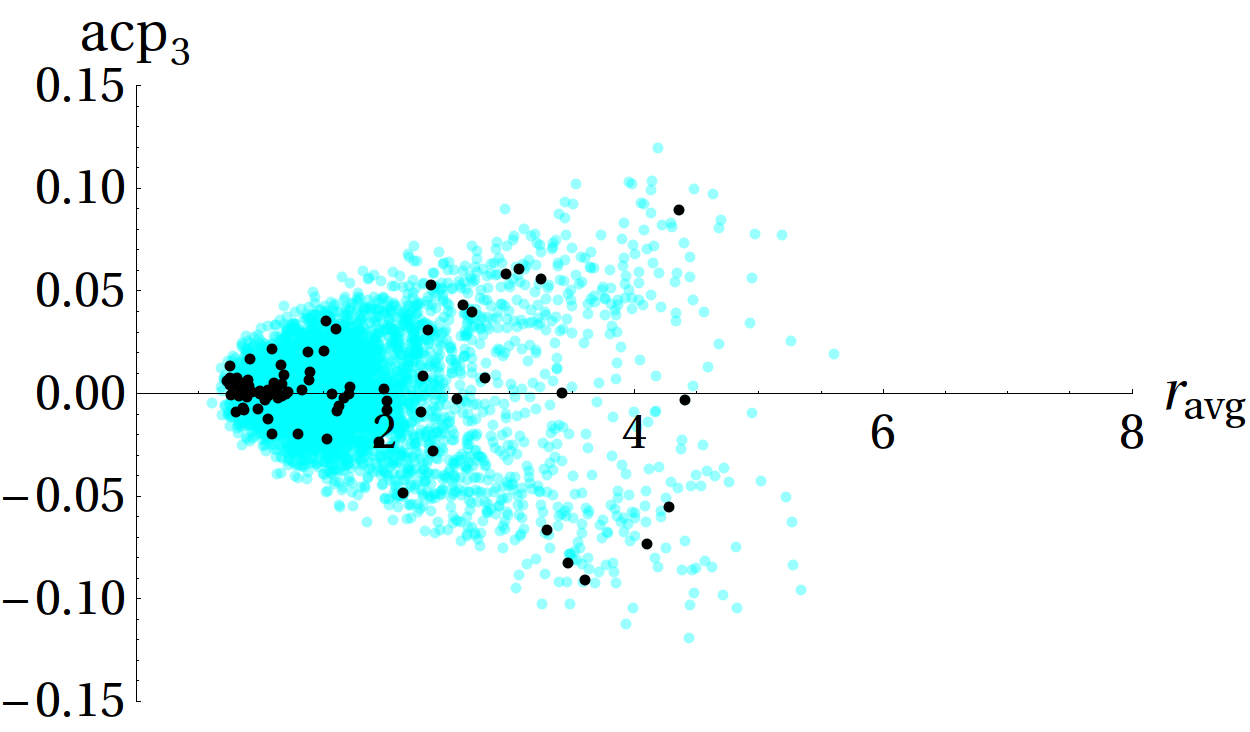}} \\[0.5em]
  \fbox{\includegraphics[width=0.48\textwidth]{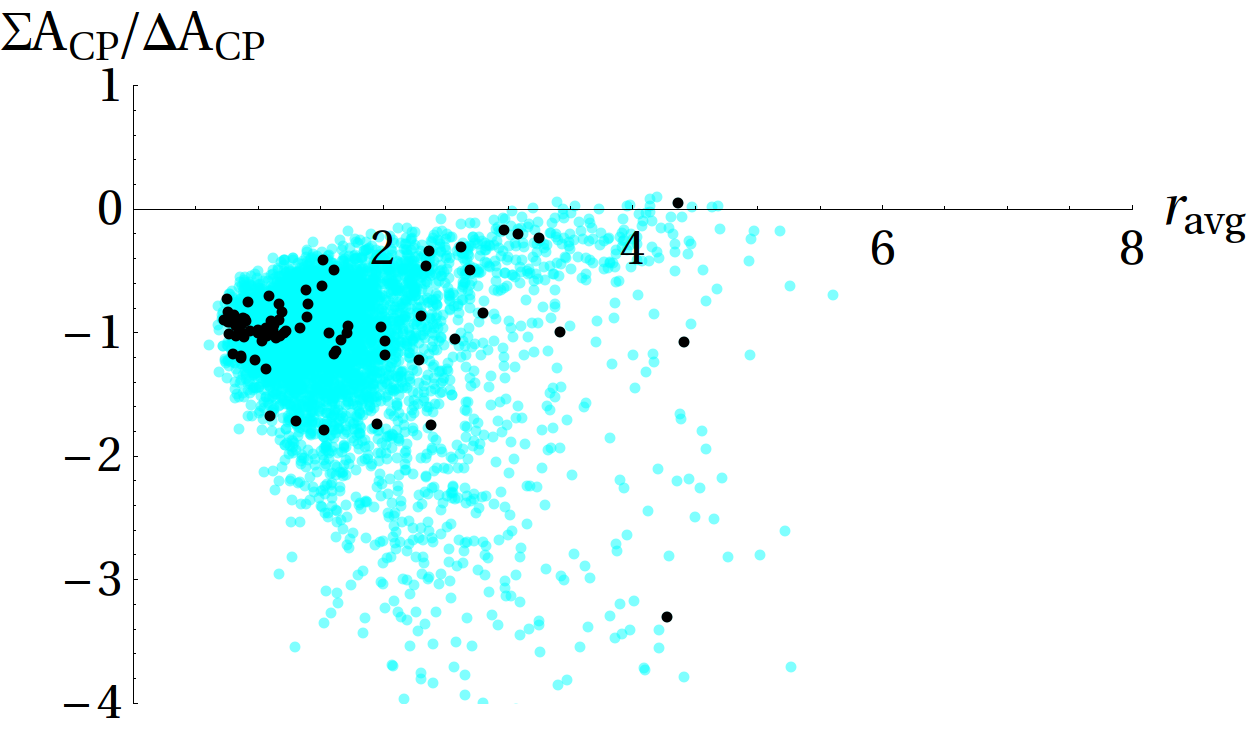}} 
 \end{center}
\caption{\label{fig:acps} Direct CP asymmetries as a function of $\rm r_{avg}$ 
 \eqref{eq:ravg} from
 a fit of W-spin parametrization to 4 observables in non-leptonic $B^0$ decays.
 Plot conventions as in fig.~\ref{fig:Uspin1}.}
\end{figure}

Despite the fact that we might expect W-spin to be badly broken, the parametrization 
is sufficient to reproduce the available experimental data. 
A comment is in order about the fine-tuning in the amplitude parameters which is
necessary to suppress the BR for the decay  $\bar B^0 \to K^- \pi^+$ 
with respect to $\bar B^0 \to D_s^- D^+$. To quantify this effect, we have determined 
the box-counting dimension (bcd) of the fitted parameter space, following \cite{Feldmann:2010iu}.
We found a bcd  which is indeed somewhat smaller than the naive
dimension, $4$ (refering to 8 fit parameters minus 4 experimental constraints). 
While each individual contributing amplitude is basically unconstrained in the fit, it turns out that 
the result for the average
\begin{align*}
 {\rm r_{avg}} = \sqrt{\frac{r^2+r_1^2+r_0^2}{3}} \,.
\label{eq:ravg}
\end{align*}
is bounded from above and, at the same time, again sets the order of
magnitude for the potential direct CP asymmetries.
This is illustrated in fig.~\ref{fig:acps}, 
where we show results for the  direct CP asymmetry
\begin{align}
{\rm acp_3} & = A_{\rm CP}^{\rm dir}[\bar B^0\to  D_s^- D^+] 
\qquad \mbox{(not yet measured)} \,,
\end{align}
and 
for the  ratio $\frac{\rm acp_3+acp_2}{\rm acp_3-acp_2}$
as a function of $\rm r_{\rm avg}$.
In the approximation of universal strong phases in $W=0$ ($W=1$) amplitudes,
the latter would be given by $\frac{1-\rm br_3/br_2 }{1+\rm br_3/br_2} \approx -1$.
The following observations can be made:
\begin{itemize}
 \item The points generated by the fit satisfy 
\begin{align}
       0.6 \leq {\rm r_{avg}}\leq 6.0 \qquad & \mbox{(for $\chi^2\leq 6)$,} \nonumber\\[0.2em]
       0.8 \leq {\rm r_{avg}}\leq 5.3 \qquad & \mbox{(for $\chi^2\leq 1)$.} 
\end{align}
 \item The maximal size of the (yet unmeasured) direct CP asymmetry $|{\rm acp_3}|$ in $\bar B^0 \to D_s^- D^+$
  is given by $\sim 12\%$ for $\chi^2<6$ (restricting ourselves to values of  $\chi^2< 1$
       we would get slightly reduced values $\sim 8-10\%$). Of course, smaller values for $\rm acp_3$,
       as expected in the factorization approximation, are not excluded in our approach. Experimental
       measurement of the CP asymmetries in $\bar B^0 \to D_s^- D^+$ would be very useful in this respect.
\end{itemize}
Of course, we should keep in mind that this picture could change when higher-order
W-spin breaking effects are taken into account. Still, we find this somewhat
academic exercise useful in getting an idea about an upper bound on the generic amplitude
ratios in non-leptonic $B^0$ and $D^0$ decays. Apparently, even for a badly broken
symmetry like W-spin, the amplitude ratios in the considered
$B^0$ decays do not exceed a value of $5-6$ or so.
Naively translated to the $D^0$ case, this implies that the amplitude ratios needed
to explain the central value of $\Delta A_{\rm CP}^{\rm dir}$ in the SM are at the upper
limit of the acceptable (or expectable) range.

\clearpage

\section{Conclusions}

Let us summarize our main results:
\begin{itemize}
 \item The ratios of BRs in $D^0 \to \pi^+\pi^-,\pi^+K^-,K^+K^-$ require ${\cal O}(1)$
       U-spin violating effects. The required magnitude of U-spin violation can be understood
       (but not unambiguously predicted) from long-distance strong-interaction effects. 

 \item The relative strong phase between the
       U-spin symmetric and U-spin violating contributions has to be large, too, which points towards
       essential non-perturbative hadronic rescattering effects. This is not surprising from experience
       on dealing with $D$-meson decays.

 \item As a consequence of U-spin violation, the direct CP asymmetries in $D^0 \to \pi^+\pi^-$ and $D^0 \to K^+K^-$
       are no longer related to be equal in magnitude and opposite in sign 
       (neither in the SM, nor in NP extensions).

 \item Within the SM, the hadronic matrix elements of  $c \to u q\bar q$ operators with
       highly Cabibbo-suppressed CKM factors should be enhanced 
       compared to the leading operators by a factor of $(3-5)$ in order to yield the observed central values 
       for the direct CP asymmetries. Although we lack a comprehensive dynamical model to generate such an enhancement, 
       there is no good reason to exclude the possibility of such numerical factors (the chiral enhancement of
       certain penguin operators in non-leptonic $B$-meson decays is a well-known example; indeed, in $K$ decays
       such enhancements are even more pronounced). As a toy example, we have
       also studied the breaking of W-spin ($u\leftrightarrow c$) in non-leptonic $B^0$ decays, which is found to exhibit
       a very similar pattern of amplitude ratios and strong phases.

 \item Due to the presence of U-spin violation, NP contributions to the measured
       $\Delta A_{\rm CP}^{\rm dir}$ are possible with both, 
       NP operators having $U=0$ or $U=1$.
 
 \item Considering the specific model of a SM extension by a fourth generation, we stress that large 
       \emph{parametric} enhancement of direct CP asymmetries (i.e.\ with less suppression in terms
        of the Wolfenstein expansion) in charm decays
       are not possible as a consequence of the tight constraints on the 4G mixing angles and CP phases from kaon and $B$-meson observables.
       Still, the additional short-distance contributions of the 4G quarks to the 
       weak effective Hamiltonian allow for a numerical enhancement (but, in principle, also to a reduction) compared to the SM.

 \item From the experimental point of view, it should be worth looking into other non-leptonic $D$-meson decay modes
 which could be accessible to LHCb or Super-B factories.
   On the one hand, there can be modes like $D^+ \to \phi\pi^+$, $D_s \to \phi K^+$
 which are induced by the same operators in the weak 
   effective Hamiltonian as $D^0 \to \pi^+\pi^-,K^+K^-$, and therefore could be
expected to yield direct CP asymmetries
   of similar magnitude.  As an example, for $D^+\to \phi\pi^+$, given the BR of about $3.1 \cdot
10^{-3}$ (including the BR for the analyzing decay $\phi \to K^+ K^-$), 
with O($10^9$) $D$ mesons, a 5$\sigma$ reach for a 0.5\% asymmetry is possible.
On the other hand, one would like to constrain direct CP violation in
tree-level decay modes such as $D^+ \to \bar K^0 (\bar K^{*0}) \pi^+$, $D_s \to \phi \pi^+ (K^+)$ etc.\ in order to test against
NP contributions in charged flavour transitions.

\item Briefly, it may be worth pointing out that analogous $D^0$ decay modes into 
 light vector and pseudoscalar mesons ($PV$) can be included, specifically
 $K^{\ast \pm} K^\mp$, $\rho^\pm\pi^\mp$. It is readily seen that significant 
 U-spin violation takes place: phase-space corrected BRs 
 (similar to $\rm obs_1$ in eq.~\eqref{obs2}) give $0.29$ and $0.21$, respectively. These modes
 can therefore be used to complement our understanding of the important issues
 related to direct CP violation in charm decays.

\end{itemize}

In conclusion, we have emphasized that the breaking of U-spin symmetry between
strange- and down-quarks points towards large non-factorizable long-distance 
effects in non-leptonic $D^0$ decays with large strong phases. As a consequence,
a SM interpretation of the present data on direct CP asymmetries in $D^0\to P^+P^-$
is plausible. On the other hand, NP models with a constrained flavour
sector, like models with a fourth fermion generation, can contribute
with a similar magnitude, leading to a moderate enhancement (or reduction) compared
to the SM.

\acknowledgments

We would like to thank Martin Jung, Thomas Mannel, and Kolya Uraltsev for
inspiring discussions. We also thank Martin Jung for critical comments 
on an earlier draft of the paper and for sharing with us the result of
a related analysis \cite{Jung:2012ip} prior to publication.
The work of AS was supported in part by the US DOE Contract \#~DE-AC~02-98CH10886.

\end{document}